\documentclass[twocolumn,showpacs,preprintnumbers,amsmath,amssymb,superscriptaddress,floatfix]{revtex4}

\usepackage{graphicx}
\usepackage{dcolumn}
\usepackage{bm}

\begin{document}

\preprint{arxiv:0904.0795}

\title{Implicit spatial averaging in inversion of inelastic x-ray scattering data}%

\author{P. Abbamonte}
  \affiliation{Department of Physics and Federick Seitz Materials Research Laboratory, University of Illinois, Urbana, IL, 61801}
\author{J. P. Reed}
  \affiliation{Department of Physics and Federick Seitz Materials Research Laboratory, University of Illinois, Urbana, IL, 61801}
\author{Y. I. Joe}
  \affiliation{Department of Physics and Federick Seitz Materials Research Laboratory, University of Illinois, Urbana, IL, 61801}
\author{Yu Gan}
  \affiliation{Department of Physics and Federick Seitz Materials Research Laboratory, University of Illinois, Urbana, IL, 61801}
\author{D. Casa}
  \affiliation{Advanced Photon Source, Aronne National Laboratory, Argonne, IL, 60439}

\date{\today}

\begin{abstract}
Inelastic x-ray scattering (IXS) now a widely used technique for studying the dynamics of electrons in condensed matter.
We previously posed a solution to the phase problem for IXS [P. Abbamonte, et. al., Phys. Rev. Lett. {\bf 92}, 237401 (2004)],
that allows explicit reconstruction of the density propagator of a system.  The propagator represents, physically, 
the response of the system to an idealized, point perturbation, so provides direct,
real-time images of electron motion with attosecond time resolution and $\AA$-scale spatial resolution.
Here we show that the images generated by our procedure, as it was originally posed, are spatial averages over all source
locations.  Within an idealized, atomic-like model, we show that in most cases a simple relationship to the 
complete, un-averaged response can still be determined.  We illustrate this concept for recent IXS measurements of single
crystal graphite.

\end{abstract}

\pacs{??, ??}
\maketitle

\section{Introduction}

Attoscience is an emerging discipline that exploits remarkable achievements in HHG laser technology to study the dynamics of electrons
in atoms, molecules, and condensed matter with sub-femtosecond time resolution.\cite{bucksbaumScience,brabec,kaptmurn}
These advances have been paralleled by also remarkable achievements in technology for 
x-ray experiments, both at the source and the endstation level.  In particular, new approaches
to inelastic x-ray scattering (IXS) provide a kinematic flexibility not previously accesible to energy-loss scattering
techniques.\cite{schuelkeBook}

We previously
If the scattering cross section is sampled over a sufficiently broad range of momentum and energy,
it can be inverted to explicitly reconstruct the density-density propagator of a material.  The propagator, 
physically, represents the density induced in a medium by an idealized, point perturbation.  Inversion 
therefore  provides direct, real time images of electron dynamics, with resolutions exceeding 20 $as$ 
in time and 1 $\AA$ in space.\cite{water}  We have used this method, for example, to show that the valence exciton in LiF is a Frenkel
exciton.\cite{LiF}

This inversion method, as it was originally posed\cite{water}, has a limitation: it produces
spatially averaged images.  To understand the origin of this limitation, 
recall that the (retarded) density propagator of a system is defined as 

\begin{equation}
\chi(x_1,x_2,t)=-i/\hbar <g| \, [ \hat{n}(x_1,t) , \hat{n}(x_2,0) ] \, |g> \theta(t)
\end{equation}

\noindent
where $|g>$ represents the ground state, and $\hat{n}(x,t)$  is the density operator

\begin{equation}
\hat{n}(x,t) = \hat{\psi}^\dagger(x,t) \hat{\psi}(x,t).
\end{equation}

\noindent
Eq. 1 represents the amplitude that a point disturbance in the electron density 
at location $x_2$ at time $t=0$ will propagate to point $x_1$ at some later time $t$.  

The two spatial coordinates, $x_1$ and $x_2$, are distinct.  That is, the form of the response
depends, in general, on the precise location of the source, $x_2$ (on top of an atom versus between
atoms, for example).  The Fourier transform of the propagator is therefore a function of two momenta

\begin{equation}
\chi=\chi(k_1,k_2,\omega).
\end{equation}

Unfortunately, in scattering there is only one momentum transfer, $k$.  IXS 
experiments do not measure all parts of the response in eq. 3, but only its longitudinal 
(or ``diagonal") components \cite{note1}

\begin{equation}
\chi(k,\omega) = \chi(k,-k,\omega).
\end{equation}

\noindent
This restricts the amount of information available with the IXS imaging approach.

If the system is translationally invariant, this restriction poses no limits on the method.
In a homogeneous system the propagator depends only on the difference $x = x_1-x_2$, i.e. 

\begin{equation}
\chi(x_1,x_2,t)=\chi(x_1-x_2,t)\equiv\chi(x,t),
\end{equation}

\noindent
its Fourier transform is a function of only one momentum

\begin{equation}
\chi = \chi(k,\omega)
\end{equation}

\noindent
and IXS can provide a complete parameterization of the density propagator.  

All real systems are made of atoms, however, so are never truly homogeneous.  One can, nonetheless,
carry out the inversion procedure described in refs. \cite{water,LiF} and construct detailed 
images of electron motion.  But these images cannot be complete.  Because they are based on a subset of all
Fourier components of the response, the images must represent some kind of average of the complete response.
For IXS inversion to be useful, it is necessary to articulate exactly what kind of average this is.

In this paper we show that the images produced in an IXS inversion
correspond to the complete response, $\chi(x_1,x_2,t)$, averaged over all source locations, $x_2$. In a homogeneous
system, in which the response is independent of $x_2$, the images can be considered complete.
Even in an inhomogeneous system, however, a simple physical picture is still applicable.
We show, within a simple but quite general model, that 
the overall size of the response scales with the electron density at the source
location, $x_2$.  The average will, therefore, be dominated by source points of high electron density.
We will show that, in cases of physical interest, the images will resemble closely the full response, 
$\chi(x_1,x_2,t)$, in which $x_2$ coincides with the peak in the electron density.  
We discuss this conclusion in the context of
some recent IXS studies of single crystal graphite. 

\begin{figure}
\centering\rotatebox{0}{\includegraphics[scale=0.45]{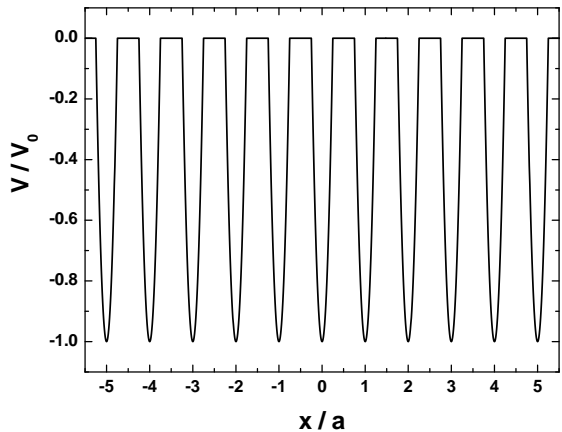}}
\caption{
Periodic array of potential wells.  For suitable choice of well depth and spacing, this system can 
be described in the tight binding approximation as a set of coupled harmonic oscillators.
}
\end{figure}

\begin{figure}
\centering\rotatebox{0}{\includegraphics[scale=0.5]{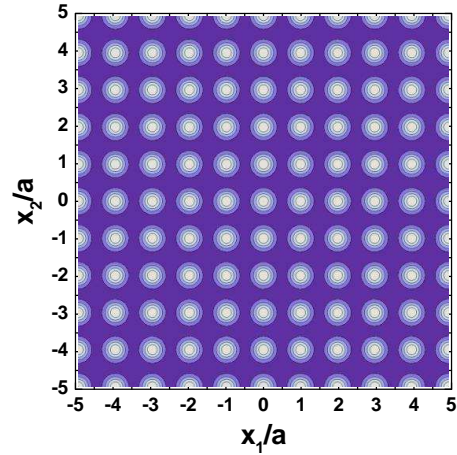}}
\caption{
Contour plot of the prefactor $\psi_0(x_1)\psi_0(x_2)$.
}
\end{figure}

\section{Inversion of IXS data}

We begin by reviewing the mechanics of an IXS inversion.  
Inelastic x-ray scattering measures the dynamic structure factor $S(k,\omega$), defined as 

\begin{equation}
S(k,\omega) = \sum_{i,f} \left | <f|\hat{n}(k)|i> \right |^2 \, P_i \, \delta(\omega-\omega_f+\omega_i)
\end{equation}

\noindent
where $P_i=e^{-\beta\omega_i}/Z$ is the Boltzmann factor ($\beta=\hbar/kT$).  The Fourier transform of $S$ 
is a space-time correlation function for the density.  
This correlation function, while interesting, is not causal so contains no direct information about
dynamics.  

$S(k,\omega)$ is, however, related to the charge propagator through the fluctuation-dissipation
theorem 

\begin{equation}
S(k,\omega)=\frac{1}{\pi} \frac{1}{1-e^{-\beta \omega}} Im[\chi(k,\omega)].
\end{equation}

\noindent
The propagator is a truly causal quantity, so provides direct information about electron dynamics.  

Inverting this cross section, to reconstruct the propagator,
requires four steps.  The first is symmetrizing to convert from $S$ to $\chi$.  
This is best done via the relationship \cite{water}

\begin{equation}
Im[\chi(k,\omega)]=-\pi \left [ S(k,\omega) - S(k,-\omega) \right ]
\end{equation}

\noindent
which does not require explicit specification of the temperature.  Next, one must
analytically continue the data.  This can be done with simple, linear interpolation\cite{water,LiF}.
Third, one performs a sine transform of this imaginary part

\begin{equation}
\chi(k,t) = \int{d\omega \, Im [\chi(k,\omega)] \, \sin{\omega t} }
\end{equation}

\noindent
where it is understood that $\chi(k,t)$ is zero for $t<0$.  Eq. (10) follows directly from the Kramers Kronig
relations and is the step in the procedure in which the phase is determined and causality is imposed.  
Finally, one performs a normal, Fourier
transform of the spatial coordinates to get $\chi(x,t)$.

This procedure has been used to image electronic motion in liquid water\cite{water},
to show that the exciton in LiF is a Frenkel exciton\cite{LiF}, and to study screening processes
in graphite and graphene\cite{uchoa}.

\begin{figure}
\centering\rotatebox{0}{\includegraphics[scale=0.45]{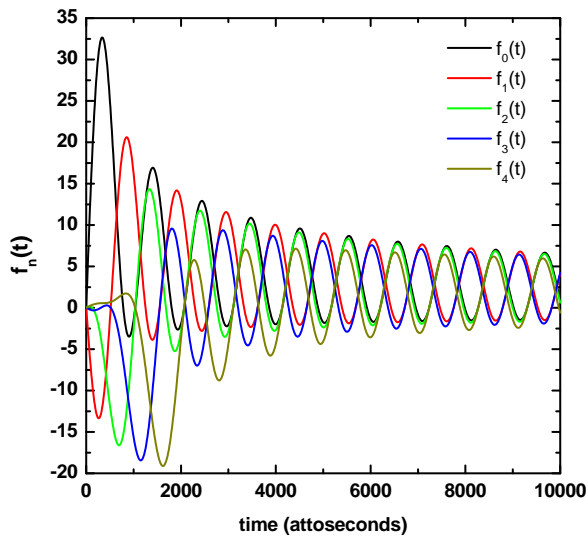}}
\caption{
Plots of the function $f_n(t)$ for values of $n$ out to fourth neighbors.  Energy rapidly spreads from the first Wigner-Seitz
cell ($n=0$) into neighboring cells.  
}
\end{figure}

\section{Implicit spatial averaging}

We now stop to ask the following question: How do we interpret the output of this procedure for
a system that does {\it not} have translational symmetry?  If we carry out the procedure described 
in Section 2, we will be left with 
a function of a single spatial variable, $\chi(x,t)$, which looks like the response of a
homogeneous system.  How should this function be interpreted?

To answer this question, we must determine an explicit relationship between this $\chi(x,t)$ 
and the complete response function $\chi(x_1,x_2,t)$.  First, we write down the Fourier 
transform

\begin{equation}
\chi(x_1,x_2)=\int{ \frac{dk_1}{2\pi}\frac{dk_2}{2\pi}\,  \chi(k_1,k_2) \, e^{ik_1x_1+ik_2x_2} },
\end{equation}

\noindent
where the time dependence has been suppressed.  To simplify the discussion we 
have assumed that our system is one-dimensional, i.e. the coordinates $x$, $x_1$, and $x_2$ are scalars.

Next, we recognize that if we insert the expression $g(k_1,k_2)=2\pi \delta(k_1+k_2)$ into the integrand, 

\begin{equation}
\chi(x_1,x_2)=\int{ \frac{dk_1}{2\pi}\frac{dk_2}{2\pi}\,  g(k_1,k_2) \, \chi(k_1,k_2) \, e^{ik_1x_1+ik_2x_2} }
\end{equation}

\noindent
eq. 11 reduces to

\begin{equation}
\chi(x_1-x_2)=\int{ \frac{dk_1}{2\pi}\,  \chi(k_1,-k_1) \, e^{ik_1(x_1-x_2)} },
\end{equation}

\noindent
which we readily recognize to be the same as eqs. 4-5.  This means that 
one can think of $\chi(x,t)$ as the full function $\chi(x_1,x_2)$ having been Fourier filtered
by the function $g(k_1,k_2)$.  

A more physical interpretation can be achieved by applying the convolution theorem.  The Fourier
transform of $g$ is given by 

\begin{equation}
g(x_1,x_2)=\delta(x_1-x_2).
\end{equation} 

\noindent
$\chi(x)$ is the Fourier transform of the product $g(k_1,k_2) \chi(k_1,k_2)$, which means it can 
be written as the convolution integral

\begin{equation}
\chi(x_1-x_2)=\int{dx'_1 dx'_2 \, g(x'_1,x'_2) \, \chi(x_1-x'_1,x_2-x'_2)}
\end{equation}

\begin{equation}
=\int{dx'_2 \, \chi(x_1-x'_2,x_2-x'_2)}.
\end{equation}

\noindent
Redefining the origin and writing in terms of $x=x_1-x_2$ we arrive at 

\begin{equation}
\chi(x)=\int{dx'_2 \, \chi(x+x'_2,x'_2)}.
\end{equation}

So, we see that the physical meaning of $\chi(x,t)$ is quite clear.  It is simply the full susceptibility,
averaged over all possible source locations, $x_2$.  No such average is done in time, however, so $\chi(x,t)$ is still 
a direct probe of electron dynamics.  Further, even in an inhomogeneous system, $\chi(x,t)$ provides direct
insight into the full response $\chi(x_1,x_2,t)$.  To illustrate why, it is useful to consider a specific case.

\begin{figure}
\centering\rotatebox{0}{\includegraphics[scale=0.45]{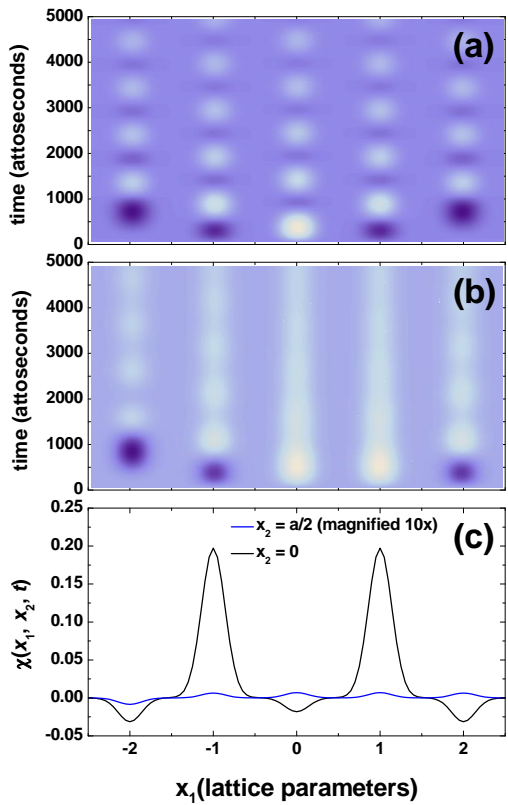}}
\caption{
(a) Contour plot of $\chi(x_1,x_2,t)$ for $x_2=0$, i.e. for the source located in the center of an ``atom".  The disturbance created
is symmetric around the source point and propagates at the average group velocity of the band. (b) Same plot, but
with $x_2=a/2$, i.e. for the source half way in between two atoms.  The disturbance is now asymmetric,
and no longer oscillates with a well-defined period.  (c) Sections through the plots in (a) and (b) at $t=1000 as$
on a linear plot, showing that the magnitude of the response in (a) is larger by $\sim 100$ times.
}
\end{figure}

\begin{figure}
\centering\rotatebox{0}{\includegraphics[scale=0.35]{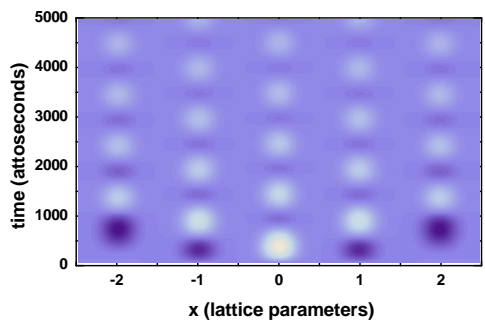}}
\caption{
$\chi(x_1,x_2,t)$ averaged over all source locations, $x_2$.  This is the image that would result by inverting IXS data.
Note the close similarity to Fig 4a.    
}
\end{figure}

\section{One electron in a periodic potential}

To illustrate the value of $\chi(x,t)$ for an inhomogeneous system, 
we consider the idealized case of a single electron traveling in a one-dimensional periodic array of 
harmonic wells,

\begin{equation}
V(x)=\sum_n v(x-na)
\end{equation}

\noindent
where each well is described by 

\begin{equation}
v(x)=k(x^2-r^2) \theta(r-|x|),
\end{equation}

\noindent
$a$ being the distance between wells and $r$ the radius of each well, as shown in Fig. 1.  For sufficiently large 
values of $k$, the ground state of an isolated well will be a Gaussian

\begin{equation}
\phi(x)=\sqrt{2\pi\sigma} \exp{-x^2/2\sigma^2}
\end{equation}

\noindent
with energy $\omega_0$.  

If the wells are weakly coupled the system can be described by a tight binding model with hopping
parameter $t_h$ (to distinguish it from the time $t$).  The states will become Bloch waves 

\begin{equation}
\psi_k(x)=\sum_n \phi(x-na) \, e^{ikna},
\end{equation}

\noindent
and the energy level will spread into a band with dispersion

\begin{equation}
\omega(k)=2t_h[1-\cos(ka)],
\end{equation}

\noindent 
where we have shifted the zero of energy to coincide with the bottom of the band, i.e. $\omega(0)=0$.

We now wish to make an explicit comparison between the full response, $\chi(x_1,x_2,t)$ for this model, and
the averaged response, $\chi(x,t)$.  These quantities can be readily computed in one-electron 
Schr\"odinger theory, but to keep the discussion general we will take a Green's function approach.  
The Fermion field operator for ths system is

\begin{equation}
\hat{\psi}(x,t) = \sum_k \, c_k \, \psi_k(x) \, e^{-i\omega(k) t}
\end{equation}

\noindent
where $c_k$ annihilates an electron with momentum $k$.  Since we have only one electron, there is no 
need to specify the spin.  In its ground state, the electron is at rest, i.e. 

\begin{equation}
|g> \, = \, c^\dagger_0 |0>
\end{equation}

Plugging these definitions into the expression for the propagator, eq. 1, we find 

\begin{widetext}
\begin{equation}
\chi(x_1,x_2,t)= \frac{8}{\pi} \left [  \psi_0(x_1) \psi_0(x_2) \right ] \cdot \left [  \sum_{n_1,n_2} \phi(x_2-n_2a) \phi(x_1-n_1a) \cdot f_{n_2-n_1}(t) \right ]
\end{equation}
\end{widetext}

\noindent
where

\begin{equation}
f_n(t)=\int_0^{\pi/a} dk \, \cos{(kna)} \, \sin{\omega(k)t} .
\end{equation}

\noindent
In general, this integral should exclude the value at $k=0$, however we may include it since $\sin{\omega(0)t}=0$.
Note that $f_n(t)=f_{-n}(t)$.

The propagator, eq. 25, consists of two factors, which can be thought of as propagators for the individual electron and
hole.  The first factor

\begin{equation}
\psi_0(x_1) \psi_0(x_2)
\end{equation}

\noindent
is independent of time and has a strong dependence on $x_2$.  We plot this factor in Fig. 2, where we have chosen
$\sigma=a/5$.  We see that this factor is strongly dependent on $x_2$, 
its maximum possible value being reached when $x_2$ resides at the
center of a well, i.e. the peak of the ground state probability density.  

We now evaluate eq. 25 and examine the properties of the full propagator.  For this purpose we evaluated 
$f_n(t)$, out to $n=5$ (fifth neighbors), using the value $t_h$ = 1 eV.  The results are shown in Fig. 3.  
The resulting $\chi(x_1,x_2,t)$ is shown as a contour plot, in Fig. 4, as a function of 
$x_1$ and $t$, for the specific values $x_2=0$ (the source in the center of a well) and $x_2=a/2$ (source 
half way between two wells).  In accordance with the physical meaning of a propagator, at $t=0$ the system 
appears to be ``struck" by an instantaneous charged perturbation, setting up a disturbance 
in the electron probability density.  As time progresses
the disturbance propagates away from the source location, dispersing through the lattice in a manner dictated by 
the dispersion of the band (eq. 22).  

When $x_2=0$, the disturbance is symmetric around $x_1=0$ and has a distinct temporal periodicity determined
by the bandwidth.  When $x_2=a/2$, on the other hand, the disturbance is asymmetric and no longer has
a clear temporal periodicity.  These two cases are distinctly different, and illustrate how
local field effects, arising from broken translational symmetry, affect the response
properties of an atomic-like system.

Most importantly, in Fig. 4c we compare, on a linear plot, the two quantities $\chi(x_1,0,t)$ and $\chi(x_1,a/2,t)$ at
a time $t=1000$ as.  Notice that the $x_2=0$ case is larger than the $x_2=a/2$ case by a factor of $\sim 200$.  
This is partly a result of the prefactor eq. 27, and has
the physical meaning that the system responds more strongly if it is ``struck" at a region of higher electron
probability density.

We now consider the averaged quantity, $\chi(x,t)$, which was determined by evaluating eq. 17, i.e. by 
integrating images like those shown 
in Fig. 4a,b for $x_2$ values between $-a/2$ to $a/2$.  The result is shown in Fig. 5.  Remarkably, it
is visually indistinguishable from Fig. 4a. 

The reason these two plots look the same is that $\chi(x_1,x_2,t)$ is  bigger when $x_2$ coincides with the center of a well.  
So the average, eq. 17, is dominated by its contribution from $x_2 \sim 0$, resulting
in a close resemblance to Fig. 4a.  In other words, the averaged quantity $\chi(x,t)$ can be thought of as the full response
$\chi(x_1,x_2,t)$ for $x_2$ residing at the peak in the electron probability density.  
This model is somewhat idealized, but this effect should occur quite generally in inhomogeneous systems,
so should aid interpretation of images generated by IXS inversion.

\begin{figure}
\centering\rotatebox{0}{\includegraphics[scale=0.45]{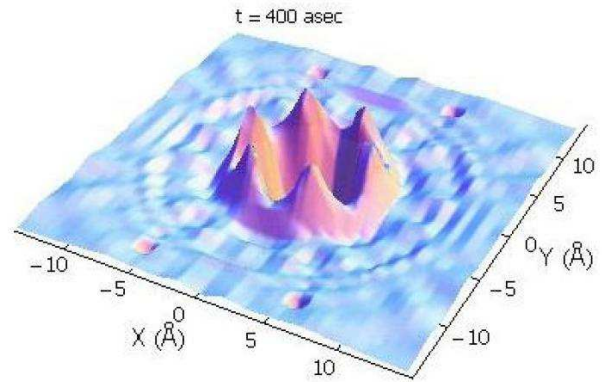}}
\caption{
$\chi({\bf r},t)$ for a real IXS experiment from a single crystal of graphite.  The six-fold symmetry is a result of the
underlying honeycomb symmetry of the graphene sheets.  This image can be thought of as a superposition of two sources,
one on the center of a C atom in each of the crystalline sub-lattices (see text).
}
\end{figure}

\section{Example: Single crystal graphite}

It is enlightening to consider a real example.  Graphite is a layered structure comprising an {\it abab}-type
stacking of graphene sheets.  The sheets have a honeycomb structure with two distinct C sites, which are
related by a translation combined with a $60^o$ rotation of the coordinates.  

Fig. 6 shows a translationally averaged $\chi({\bf r},t)$, at time $t=400 as$, acquired by inversion of IXS
data from a single crystal (not HOPG) of graphite.  The images show a pronounced six-fold symmetry, which
is result of the underlying hexagonal symmetry of the lattice.
The experiments were done at Sector 9 XOR of the Advanced Photon Source.
The details of this experiment are being published in two separate articles\cite{reed,uchoa}.  For the current purposes, 
we wish only to point out the following.

The ground state electron density of graphite is maximum at the center of a C atom.  In accordance
with conclusion of Section IV, we expect that Fig. 6 will resemble the full $\chi({\bf r}_1;{\bf r}_2,t)$,
for the specific case of ${\bf r}_2$ centered on an C atomic site.

Unfortunately, there are two C atoms in a unit cell.  They are distinct but related by a simple symmetry operation,
so should have the same electron density.  Where should one visualize the source residing in this case?  
Further, the surroundings of each C atom has a three-fold, rotational symmetry.  Why does Fig. 6 exhibit six-fold symmetry?

The point here is that, by eq. 17, the image Fig. 6 is {\it coherent} superposition of sources at the two 
distinct C locations.  Because they have the same density, they contribute equally to the average in eq. 17.  Further,
while each C atom locally has only three-fold symmetry, one is rotated $60^o$ from the other.  The superposition  
therefore exhibits six-fold symmetry, as observed in Fig. 6.  Because we know that, apart
from the $60^o$ rotation, the two C atoms are the same, one can in principle reconstruct the response of an 
isolated C atom from Fig. 6.  The important point for now is that eq. 17 provides
a useful framework for conceptualizing the meaning of inverted IXS data. 

It is worth mentioning that it
may be possible to extend IXS inversion methods to image more than just an average response.
Many years ago it was shown \cite{sw1,sw2} that by
setting up a standing wave in a material -- for example by hitting a Bragg reflection -- one can measure a sub-class
of the off-diagonal components of the susceptibility.  Under standing wave conditions, the initial
photon will be in a superposition of two different momentum eigenstates, in which case the IXS cross section will
contain contributions from  $\chi(k_1,k_2,\omega)$, where $k_1 \neq k_2$.  In principle this method can be used 
to ``anchor the problem" in space, and potentially reconstruct the complete $\chi(x_1,x_2,t)$.  Whether it is experimentally
possible to measure a sufficiently large set of these off-diagonal terms is a subject of current investigation.

In summary, we have shown that IXS imaging, as prescribed in ref. \cite{water}, provides images of the 
charge propagator, $\chi(x_1,x_2,t)$, that are averaged over all source locations, $x_2$.  
No such average is done in time, however, so this method is still a {\it direct} probe of electron dynamics in the 
attosecond regime.  The size of the propagator is much larger when $x_2$ resides in a region of high density, 
so one can think of the averaged images as corresponding to the complete response, for $x_2$ 
residing at the peak of the electron density.  Standing wave methods, in which the incident photon is placed in a 
superposition of momenta may extend the applicability of the technique.

We thank David Cahill, Gerard C. L. Wong and Robert Coridan for helpful discussions, and Xiaoqian Zhang
for a careful reading of the manuscript.
This work was funded by the Materials Sciences and Engineering Division,
Office of Basic Energy Sciences, U.S. Department of Energy under grant no. DE-FG02-07ER46459.
Use of the Advanced Photon Source was supported by DOE contract no. DE-AC02-06CH11357.  

\bibliography{TransAv}

\begin{thebibliography}{11}
\bibitem[{not()}]{note1}
\bibinfo{note}{IXS measures, of course, only the imaginary part
  $Im[\chi(k,-k,\omega)]$, but the real part can be retrieved by Kramers Kronig
  transform.}
\end{thebibliography}

\end{document}